\documentclass[runningheads]{llncs}
\usepackage{float}
\usepackage{amsmath}
\usepackage{amssymb}
\usepackage{subcaption}
\usepackage{booktabs}
\usepackage[section]{placeins}
\usepackage{multirow}

\usepackage[T1]{fontenc}
\usepackage{graphicx}
\usepackage{color}

\begin{document}
\title{Autonomous Chaotic Time Series Prediction using Physical Neuromorphic Networks}
\titlerunning{Predicting Chaotic Dynamics with Neuromorphic Networks}
%
\author{Akshaya Rajesh\inst{1}\orcidID{0009-0003-1531-4053} \and
Yinhao Xu\inst{1}\orcidID{0009-0002-8309-9650} \and
\\
Wave Ngampruetikorn\inst{1}\orcidID{0000-0001-8548-3029} \and
\\
Zdenka Kuncic\inst{1}\orcidID{0000-0001-6765-3215}}
\authorrunning{A. Rajesh et al.}
%
\institute{University of Sydney, Sydney NSW 2006, Australia \\
\email{zdenka.kuncic@sydney.edu.au}\\
}
\maketitle              
\begin{abstract}
Physical reservoir computing (PRC) with neuromorphic networks offers a promising approach to 
brain-inspired information processing, exploiting 
emergent nonlinear dynamics of physical neural networks 
as a computational resource. This study demonstrates 
fully autonomous closed-loop prediction of the 
Mackey--Glass (MG) chaotic time series using a simulated 
neuromorphic nanowire network as the physical reservoir. 
Two strategies are evaluated: the virtual node 
(VN) method, which expands the feature space by temporal 
multiplexing of reservoir states, and a non-VN approach 
that uses all physical node readouts directly without 
temporal multiplexing. Results are reported for two 
values of the MG time delay parameter, $\tau = 18$ and 
$\tau = 21$, the latter representing a more complex 
chaotic regime not previously evaluated for this class 
of physical reservoir. Over a short prediction horizon 
of $T = 100$ timesteps, the VN approach achieves 
autonomous prediction accuracies of $90.4$\% 
and $89.7$\% at $\tau = 18$ and $\tau = 21$, 
respectively, while the non-VN approach achieves 
$81.5$\% and $76.2$\%. Long-horizon 
analysis over $T = 500$ timesteps shows that both 
approaches reproduce the qualitative attractor structure 
and dominant spectral content of the true MG signal, 
with trajectories remaining bounded throughout. These 
results suggest that the intrinsic dynamics of 
neuromorphic nanowire networks are sufficient to support 
meaningful autonomous chaotic time series prediction 
without virtual node augmentation, and that performance 
may improve further as physical network sizes scale to 
the millions of nodes achievable in hardware. As this 
study uses simulated networks, extrapolation to 
physically fabricated large-scale arrays remains to be 
validated experimentally.

\end{abstract}

\section{Introduction}

Time series forecasting is the ability to predict the future state of a dynamical system given its past state history. When the system dynamics are chaotic (i.e. high nonlinearity and sensitivity to initial conditions), time series prediction is particularly challenging. Statistical forecasting techniques for chaotic time series prediction have expanded from traditional methods to include machine learning based regression techniques and deep learning models based on recurrent neural networks (RNNs) (see e.g. \cite{gilpinModel2023} for an overview).

The best performance in chaotic time series forecasting with RNNs is achieved with the relatively ``shallow'' architecture used in reservoir computing (RC), where a single RNN projects the time-series data into a higher-dimensional feature space \cite{pathakModelFree2018}. Crucially, the RNN itself is not trained; its weights are typically randomized and fixed. The node states are used as features to train a single linear output layer, so model training is relatively lightweight \cite{lukoseviciusReservoir2009}.

A new paradigm in RC is physical reservoir computing (PRC), where the reservoir (the RNN) is a physical system whose intrinsic dynamics effectively perform the nonlinear transformation of input signals that would otherwise be transformed by a mathematical function (e.g. tanh) in RC \cite{tanakaRecent2019}. A particularly interesting example of PRC is with physical RNNs comprised of nanowires that naturally self-assemble into a complex, recurrent architecture that exhibits remarkable similarity to biological neural network topologies \cite{loeffler2020topological}. Moreover, these physical neuromorphic networks exhibit  brain-like dynamics under electrical stimulation \cite{kuncicNeuromorphic2021,hochstetterAvalanches2021}.

Previous studies have demonstrated various PRC tasks with neuromorphic nanowire networks, including image and audio classification \cite{Lilak2021,Milano_2022,zhu2023online,Kotooka2024}, sequence memory \cite{zhu2023online} and waveform regression \cite{sillin2013theoretical,hochstetterAvalanches2021,zhuInformation2021,Loeffler2021}, as well as chaotic time series prediction \cite{Zhu_transfer2020,Fu2020,milano2023mackey,baccettiErgodicity2024,Xu_ijcnn2025,xu2025}. A common chaotic time series prediction benchmark task for RC and PRC is forecasting the Mackey--Glass (MG) dynamical system \cite{MackeyGlass1977}, which is a nonlinear time delay differential equation that becomes chaotic above a threshold value of its time delay parameter \cite{jaegerHarnessing2004}. Previous PRC studies with nanowire networks demonstrated MG prediction when the system becomes marginally chaotic and although one study \cite{Zhu_transfer2020} showed how transfer learning could be applied to learn the MG system with more chaotic dynamics, the prediction was performed non-autonomously (i.e. with regular updates -- see also \cite{Fu2020}). While autonomous prediction of the MG system was demonstrated in Ref.~\cite{milano2023mackey}, the PRC approach relied on the virtual nodes method to expand the nanowire network reservoir states (see e.g. \cite{Kitano_2026} for a recent overview of this method for PRC).

This study demonstrates autonomous prediction of the MG 
system using a nanowire network as the physical reservoir, 
for two values of the time delay parameter corresponding 
to increasingly complex chaotic dynamics. Results are 
compared between two readout strategies: the standard 
virtual node (VN) method and a non-VN approach that 
relies solely on the intrinsic spatial diversity of 
network node readouts. Sec.~\ref{sec:methods} outlines 
the methods and Sec.~\ref{sec:results} presents and 
discusses the results.

\section{Methods}
\label{sec:methods}

Autonomous prediction of the MG chaotic time series was performed using PRC with a simulated neuromorphic nanowire network as the reservoir.

\subsection{Neuromorphic Network}

\begin{figure}[htbp]
\includegraphics[width=\textwidth]{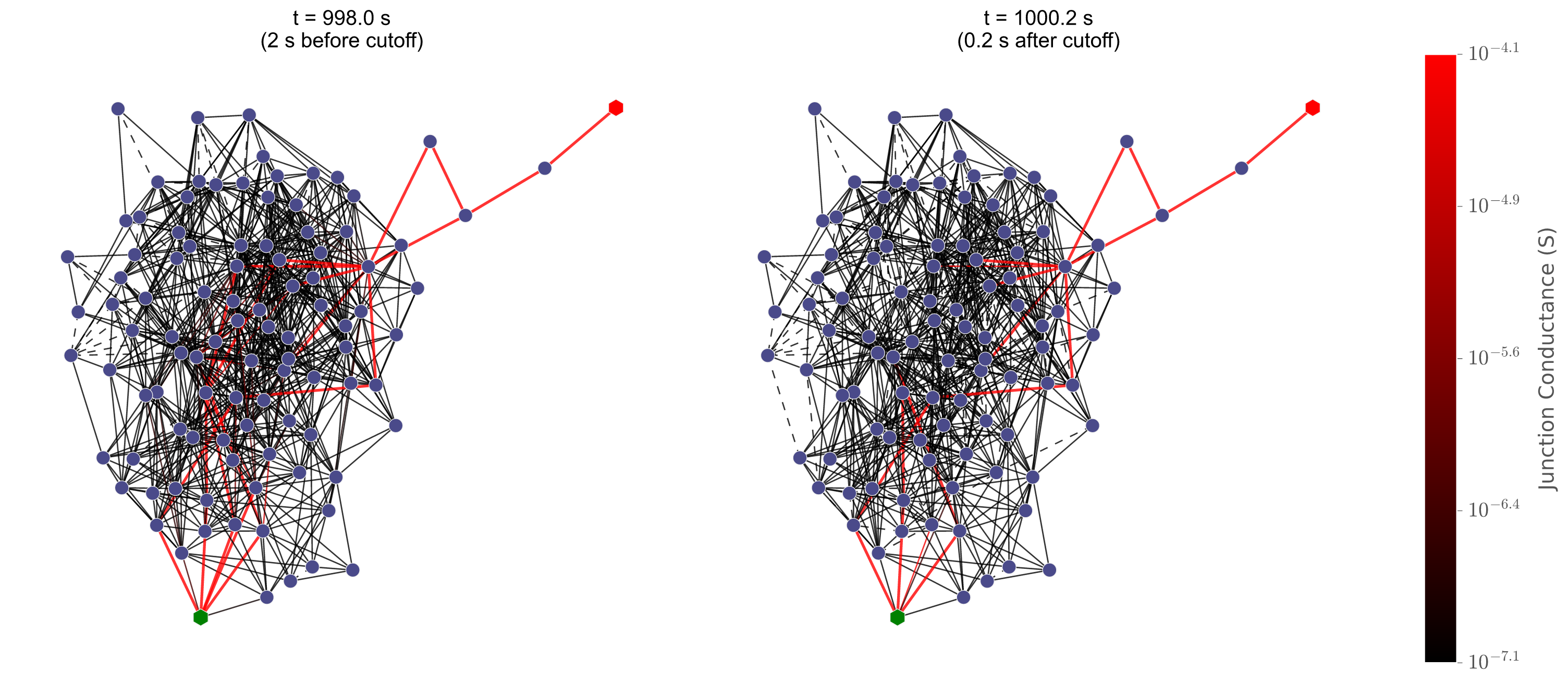}
\caption{Graph visualisation of a simulated neuromorphic nanowire 
network (100 nodes shown for clarity) at two snapshots: 2\,s before (left) and 0.2\,s 
after (right) the input signal drops below $V_{\text{reset}}$. 
Colourbar indicates junction conductance, with dashed edges indicating junctions below 
the visualisation threshold. Red and green nodes mark the input 
and ground.
}
\label{fig:networkgraph}
\end{figure}

The neuromorphic network used in this study is based on a physically motivated model of self-assembled nanowire networks that natively produce brain-like dynamics under electrical stimulation \cite{kuncicNeuromorphic2021}. The model is described in detail elsewhere
\cite{kuncicNeuromorphic2020a,hochstetterAvalanches2021,zhuInformation2021,xu2025}. The neuromorphic network is abstracted as a graph 
in which nodes represent nanowires and edges represent 
nanowire-nanowire cross-point junctions (see Fig.~\ref{fig:networkgraph}). Under an electrical bias, the junctions exhibit resistive switching memory, henceforth referred to as memristive junctions. Kirchhoff's circuit laws for electric charge and energy conservation are enforced at every timestep to solve for node voltages as a function of 
the applied input voltage signal. The junction edges represent dynamic conductance-based weights that
evolve according to the following memristor equation of state,
\begin{equation}
\dot{x}_i = 
\begin{cases}
(|v_i| - V_{\text{set}})\,\text{sgn}(v_i) & |v_i| > V_{\text{set}} \\
0 & V_{\text{reset}} \leq |v_i| \leq V_{\text{set}} \\
b(|v_i| - V_{\text{reset}})\,\text{sgn}(x_i) & |v_i| < V_{\text{reset}}
\end{cases}
\end{equation}
\noindent where $x_i(t)$ is the internal state variable of junction $i$ with voltage $v_i (t)$,  $V_\text{set}$ and $V_\text{reset}$ are voltage thresholds governing growth and decay, respectively, and $b$ determines decay rate.
Junction conductance $g_i(x_i)$ 
is a nonlinear function of $x_i$, spanning approximately three orders of 
magnitude between its off and on states \cite{hochstetterAvalanches2021}. Unless 
otherwise stated, all results presented in this study are obtained using simulated networks with 500 nodes and 9,905 edges.

\begin{figure}[htbp]
\includegraphics[width=\textwidth]{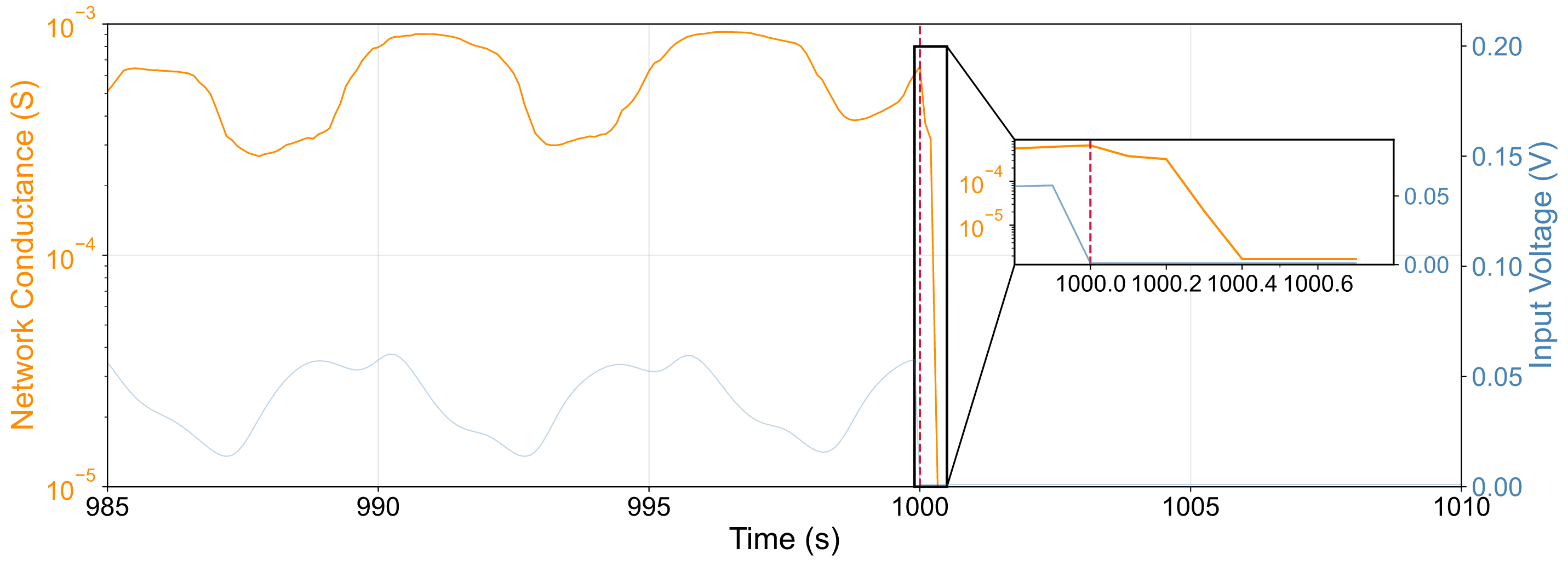}
\caption{Network conductance (orange) after delivering an input voltage MG signal (blue) to a neuromorphic nanowire network for 1000\,s (only the final 15\,s and following 10\,s is shown). Inset is a zoom-in showing conductance decay over 0.4\,s after input drops to below the $V_\text{reset}$ threshold.}
\label{fig:decay}
\end{figure}

Fig.~\ref{fig:decay} shows an example of the short-term memory response of the nanowire network model. 
Following 1000\,s of input signal, after which the voltage 
drops to below $V_\text{reset}$, the network conductance decays over a period of $\approx 0.4$\,s. 
This confirms that the network retains a short-term  memory of its input history. The nonlinear input signal used in Fig.~\ref{fig:decay} is the MG chaotic time series with $\tau = 18$.

\FloatBarrier
\subsection{Mackey-Glass Time Series}

The MG nonlinear time series is generated using the discrete-time recurrence
\begin{equation}
y[n+1] = y[n] - \gamma y[n] + \beta \frac{y[n-\tau]}{1 + y[n-\tau]^{10}},
\label{eq:mg}
\end{equation}
\noindent with parameters $\beta = 0.2$ and $\gamma = 0.1$. This is the 
Euler discretisation of the MG delay differential equation 
\cite{MackeyGlass1977} with unit time step and time delay parameter $\tau$, and is the standard formulation 
used in RC benchmarks \cite{jaegerHarnessing2004}. The MG time series becomes marginally chaotic when $\tau \approx 18$, which is the value commonly used in prediction tasks. This study also considers $\tau = 21$, which can produce more complex dynamics \cite{tarigo_MG_2022}. The time series is initialized with a fixed sequence of 18 
values and is mean-centred prior to use. A washout period of 100 steps is 
discarded to eliminate transient dependence on initial conditions.

\subsection{Physical Reservoir Computing}

Autonomous prediction of the MG time series is implemented within the PRC framework \cite{tanakaRecent2019,Kan2021,Milano2023strategies,Liang_2024}, 
in which the nanowire network serves as a high-dimensional nonlinear dynamical system. An input scalar $u(t)$ is scaled by a voltage amplitude parameter 
$\alpha$ and delivered to a subset of input nodes, $N_i$, while voltage 
readouts from a disjoint subset of $N_r$ nodes are used as the 
feature vectors presented to a linear output layer. Only the output layer weights are trained (see Sec.~\ref{sec:training}). For all simulations, $N_i = 25$ input nodes are selected and used to send the same MG input signal.

Each readout node contributes a single voltage value per timestep, yielding a feature vector of dimension $N_r$.
The virtual nodes (VN) method \cite{appeltant2011information} is used to expand the feature space dimensionality by sampling the reservoir state (i.e. nodes) at equally spaced sub-timestep intervals 
within each input period and results are compared to that obtained without relying on virtual node timeplexing, using only the network node readouts \cite{Kitano_2026}.
The 20 virtual nodes are sampled at equally spaced sub-intervals $\theta = T/20$ within each input period of duration $T$, where $T$ was also optimized per evaluation shift within $[0.05,\, 1]$~s; the physical simulation timestep was set equal to this spacing ($dt = T/20$), giving one simulated sample per virtual node. For the representative trajectories shown in Sec.~\ref{sec:results}, this corresponds to $\theta = dt = 2.875$~ms ($\tau = 18$) and $\theta = dt = 5.691$~ms ($\tau = 21$).

The VN and non-VN methods are described in Sec.~\ref{sec:training} below.

\subsection{Training and Autonomous Prediction}
\label{sec:training}

Following 
Ref.~\cite{xu2025}, the predicted signal is:

\begin{equation}
\hat{y}(t) = \mathbf{W}_{\text{out}}\, \mathbf{r}(t) 
+ \mathbf{u}(t),
\label{eq:skip}
\end{equation}

\noindent where $\mathbf{W}_{\text{out}} \in \mathbb{R}^{1 \times N_r}$ is the output weight matrix, $\mathbf{r}(t)$ is the reservoir state 
vector at the current time $t$ and $\mathbf{u}(t)$ is the current input signal, which serves as a skip connection. 
This means the network is used to learn the 
residual $\Delta \mathbf{y}(t)$ rather than the signal itself, which has been shown to stabilize 
autonomous prediction of chaotic dynamical systems~\cite{xu2025}. The output weight matrix 
 is trained using ridge regression,

\begin{equation}
\mathbf{W}_{\text{out}} = \mathbf{Y} \mathbf{R}^{\top} 
\left( \mathbf{R} \mathbf{R}^{\top} + \lambda \mathbf{I} 
\right)^{-1},
\label{eq:ridge}
\end{equation}

\noindent where $\mathbf{R}\in\mathbb{R}^{d\times N}$ is the matrix of reservoir 
feature vectors collected over the training period, 
and $\mathbf{Y}\in\mathbb{R}^{1 \times N}$ contains the corresponding target values, 
and $\lambda$ is the regularisation coefficient. 
For the VN approach, 20 virtual nodes are used per 
readout node. Correspondingly, $\mathbf{W}_{\text{out}} \in \mathbf{R}^{1 \times 20 N_{r,\text{VN}}}$ for the VN approach, where $N_{r,\text{VN}}$ is the number of readout nodes selected for virtual-node expansion, while $\mathbf{W}_{\text{out}} \in \mathbf{R}^{1 \times N_r}$ as defined above applies to the non-VN approach.
For the non-VN approach, all readout node voltages are used directly as  
feature vectors, with no temporal multiplexing. 
During training, $\mathbf{u}(t) = \mathbf{y} (t - \Delta t)$, i.e. the MG signal at the previous timestep.
For closed-loop autonomous prediction, the previous estimate is used as input, i.e. $\mathbf{u}(t) = \hat{\mathbf y} (t - \Delta t)$,
with no access to the true MG signal during the autonomous phase.

Hyperparameters are selected via Bayesian optimization 
using Optuna~\cite{akiba2019optunanextgenerationhyperparameteroptimization}. For the VN 
approach, the search space comprises the regularization 
coefficient $\lambda \in [10^{-6},\, 1]$ (log scale) 
and input amplitude $\alpha \in [0.005,\, 2.0]$. For 
the non-VN approach, the search space comprises 
$\lambda \in [10^{-6},\, 10^{-1}]$ (log scale), $\alpha \in [0.04,\, 0.7]$, 
and 
training duration $t_{\text{train}} \in [20000,\, 40000]$ 
timesteps. Optimization is performed independently for 
each value of $\tau$: 18 and 21. The non-VN representative trajectory shown in Sec.~\ref{sec:results} uses $\alpha = 0.094$, $\lambda = 2.1\times10^{-4}$, $t_{\text{train}} = 32616$ ($\tau = 18$) and $\alpha = 0.284$, $\lambda = 3.1\times10^{-8}$, $t_{\text{train}} = 23634$ ($\tau = 21$).

For the VN approach, training duration was fixed at $t_{\text{train}} = 1000$ timesteps, unlike the non-VN approach where $t_{\text{train}}$ was included in the search space. The number of readout nodes $N_{r,\text{VN}}$ used for virtual-node expansion was also Bayesian-optimized per shift, with a search space of  $[10, 497]$ candidate nodes drawn from the 475 electrodes available (500 total nodes minus 25 input nodes). Each readout node in the VN approach expanded to 20 virtual nodes. The representative trajectory shown in Sec.~\ref{sec:results} uses $\alpha = 3.094$, $\lambda = 7.29\times10^{-5}$, $N_{r,\text{VN}} = 452$ ($\tau = 18$) and $\alpha = 1.794$, $\lambda = 1.99\times10^{-3}$, $N_{r,\text{VN}} = 470$ ($\tau = 21$).

\subsection{Performance Metrics}

Prediction accuracy is quantified using the normalised root mean 
squared error (NRMSE), computed over signals rescaled to $[0,1]$:

\begin{equation}
    \text{NRMSE} = \frac{1}{y_{\max} - y_{\min}} 
    \sqrt{ \frac{1}{T} \sum_{t=1}^{T} 
    \left( \hat{y}(t) - y(t) \right)^2 },
    \label{eq:nrmse}
\end{equation}

\noindent where $\hat{y}(t)$ and $y(t)$ are the predicted and 
true signals respectively, $T$ is the number of prediction 
timesteps, and $y_{\max} - y_{\min}$ is the range of the true 
signal. Dividing by the signal range renders the metric 
scale-invariant, enabling direct comparison across configurations 
with different input amplitudes and MG delay parameters $\tau$. 
Accuracy is calculated as $1 - \text{NRMSE}$ and reported as mean $\pm$ standard deviation (s.d.) 
across 10 independent trials with different random network 
realizations.

For assessing long-horizon prediction ($T = 500$ timesteps), 
the predicted and true MG trajectories are compared 
qualitatively in phase space, and quantitatively via their 
normalized power spectral densities (PSDs).

\section{Results and Discussion}
\label{sec:results}

The performance of nanowire network PRC on 
closed-loop autonomous MG prediction is evaluated 
for $\tau = 18$ and $\tau = 21$. Two readout strategies are 
compared: the virtual node (VN) approach, which augments a 
Bayesian-optimized subset of readout nodes with 20 virtual 
nodes each, and the non-virtual node (non-VN) approach, which 
uses all available readout nodes directly without temporal 
multiplexing. Short-term prediction accuracy is assessed over 
$T = 100$ timesteps, while long-horizon forecasting is analyzed 
over $T = 500$ timesteps in phase and frequency space.

\subsection{Short-Term Autonomous Prediction}
\label{sec:short_term}

\begin{table}[htbp]
\centering
\caption{Autonomous prediction accuracy 
($1 - \text{NRMSE} \pm 1$\,s.d.) and corresponding NRMSE for VN and non-VN 
readout strategies at $\tau = 18$ and $\tau = 21$, 
averaged over 10 independent trials for $T = 100$ steps.}
\label{tab:combined_accuracy}
\renewcommand{\arraystretch}{1.6}
\setlength{\tabcolsep}{6pt}
\begin{tabular}{@{}c l cc cc@{}}
\toprule
& & \multicolumn{2}{c}{\textbf{Accuracy}} & \multicolumn{2}{c}{\textbf{NRMSE}} \\
\cmidrule(l){3-4}\cmidrule(l){5-6}
$\boldsymbol{\tau}$ & \textbf{Method} 
& \textbf{Train} & \textbf{Autonomous} & \textbf{Train} & \textbf{Autonomous} \\
\midrule
\multirow{2}{*}{18} 
  & VN    
  & $0.9948 \pm 0.0031$ 
  & $0.9045 \pm 0.0386$ 
  & $0.0052 \pm 0.0031$
  & $0.0955 \pm 0.0386$ \\[4pt]
  & Non-VN 
  & $0.9655 \pm 0.0012$
  & $0.8153 \pm 0.0592$
  & $0.0345 \pm 0.0012$
  & $0.1847 \pm 0.0592$ \\
\addlinespace[6pt]
\multirow{2}{*}{21} 
  & VN    
  & $0.9943 \pm 0.0025$ 
  & $0.8975 \pm 0.0332$
  & $0.0057 \pm 0.0025$
  & $0.1025 \pm 0.0332$ \\[4pt]
  & Non-VN 
  & $0.9530 \pm 0.0043$
  & $0.7619 \pm 0.0686$
  & $0.0470 \pm 0.0043$
  & $0.2381 \pm 0.0686$ \\
\bottomrule
\end{tabular}
\end{table}

Table~\ref{tab:combined_accuracy} reports prediction accuracies 
for the VN and non-VN approaches. For $\tau = 18$, the VN approach achieves an autonomous prediction 
accuracy of $0.904 \pm 0.039$ over $T=100$ steps, which is consistent with 
results reported in previous nanowire network PRC studies on MG prediction \cite{Milano_2022,milano2023mackey}.
Another similar study \cite{Fu2020} achieved a higher accuracy of $\approx 98$\% over $T=50$ steps, but only by using 50 virtual nodes and applying regular teacher-signal updates, so prediction was not fully autonomous.
Similarly, Ref.~\cite{Zhu_transfer2020} used the past states of readout nodes to achieve $\approx 74$\% over $T=20$ steps for $\tau =18$.
The non-VN approach
achieves an autonomous prediction accuracy of 
$0.815 \pm 0.059$ for $\tau = 18$. While 9\% lower than the VN 
approach, the reduction is moderate, and the larger 
standard deviation reflects greater sensitivity to 
network initialization when no readout optimization is 
applied.

For $\tau = 21$, the VN approach achieves an autoprediction accuracy of $0.897 \pm 0.033$, representing a modest reduction relative to $\tau = 18$. This is consistent with the increased memory 
depth required at longer delays, which places greater 
demand on the network's short-term fading memory.
The standard deviations across trials are 
relatively low at both $\tau$ values, indicating that 
performance is broadly consistent across different network 
initializations.
The non-VN approach achieves an autoprediction accuracy of $0.762 \pm 0.069$ for $\tau = 21$.
While this is 13\% lower than the corresponding accuracy for the VN approach, 
it is achieved without any temporal 
multiplexing, node selection, or virtual node 
construction, suggesting that the intrinsic 
diversity of the 500-node network readout alone is 
sufficient to support meaningful autonomous prediction.

\begin{figure}[htbp]
    \centering
    \begin{subfigure}{\textwidth}
        \centering
        \includegraphics[width=0.75\textwidth]{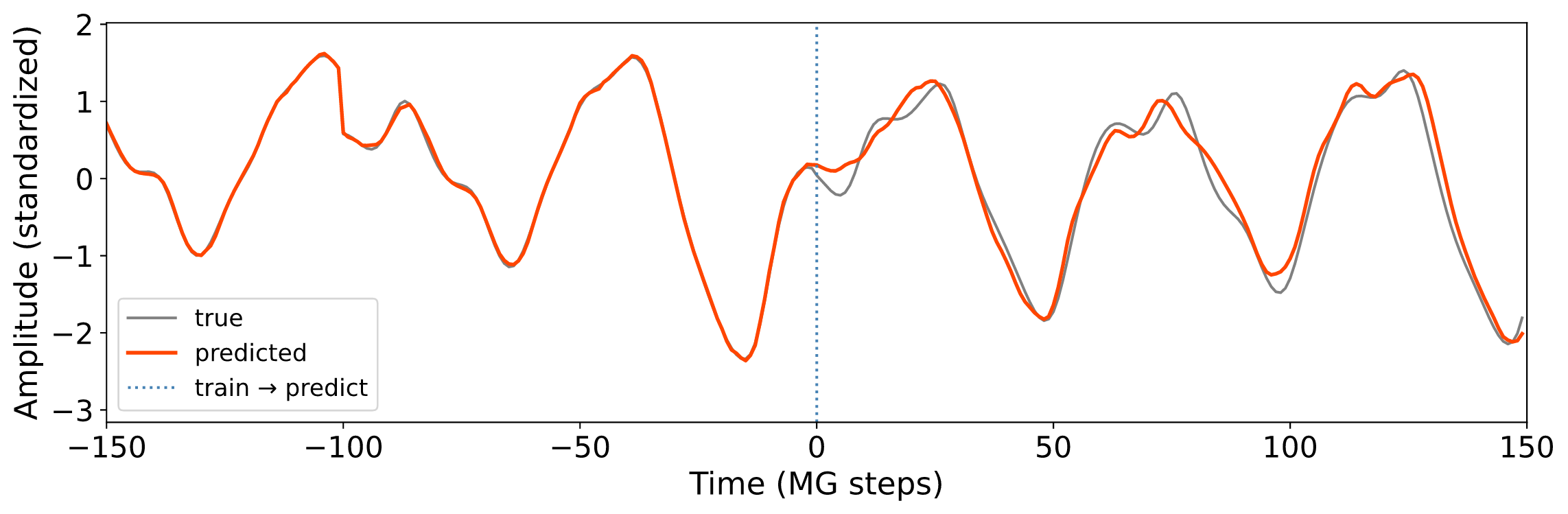}
        \caption{$\tau = 18$}
        \label{fig:vn_time_18}
    \end{subfigure}
    
    \vspace{1em}
    
    \begin{subfigure}{\textwidth}
        \centering
        \includegraphics[width=0.75\textwidth]{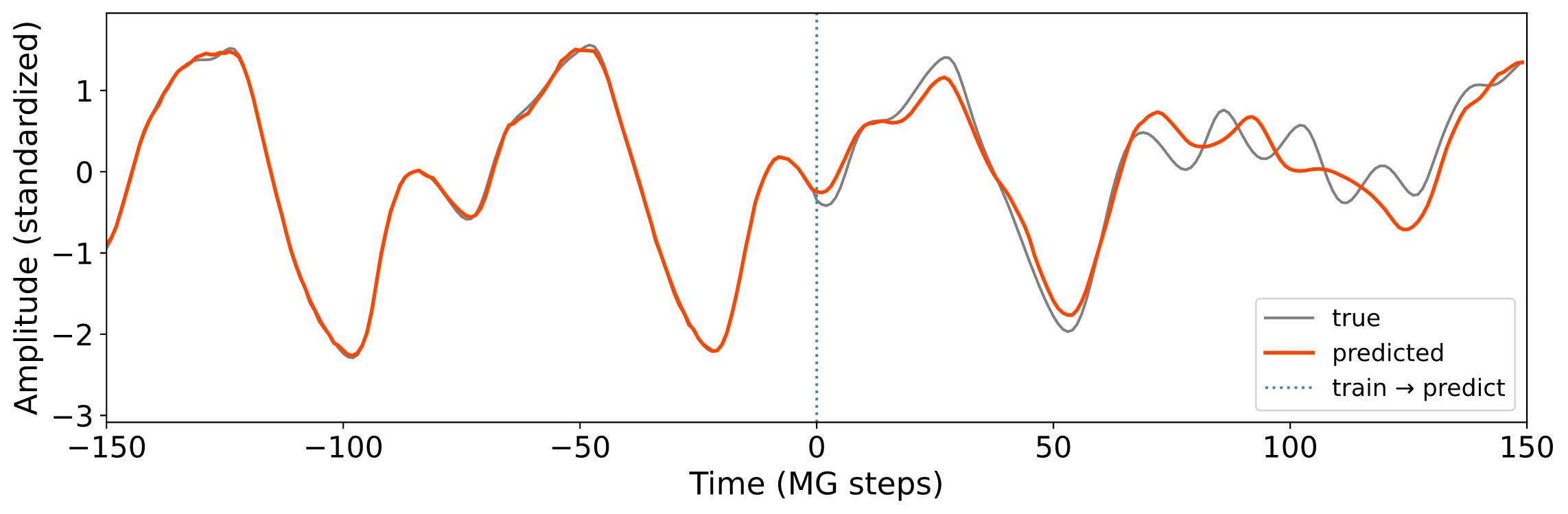}
        \caption{$\tau = 21$}
        \label{fig:vn_time_21}
    \end{subfigure}
    
    \caption{Autonomous prediction of the Mackey-Glass 
    time series over using the VN method. Top: 
 $\tau = 18$; bottom: 
    $\tau = 21$. The vertical dashed line marks the 
    transition from teacher-forced training to closed-loop 
    autonomous prediction.}
    \label{fig:vn_timeseries}
\end{figure}

Figure~\ref{fig:vn_timeseries} shows representative 
time-series predictions for the VN approach. In both 
panels, the predicted signal tracks the true trajectory 
closely in the immediate post-training window, with phase 
drift accumulating at later timesteps, a behaviour 
expected for chaotic systems, where small prediction 
errors grow exponentially over time. The $\tau = 21$ case shows 
slightly earlier phase decoherence, consistent with the 
greater sensitivity to initial conditions at this delay 
value.

\begin{figure}[htbp]
    \centering
    \begin{subfigure}{\textwidth}
        \centering
        \includegraphics[width=0.75\textwidth]{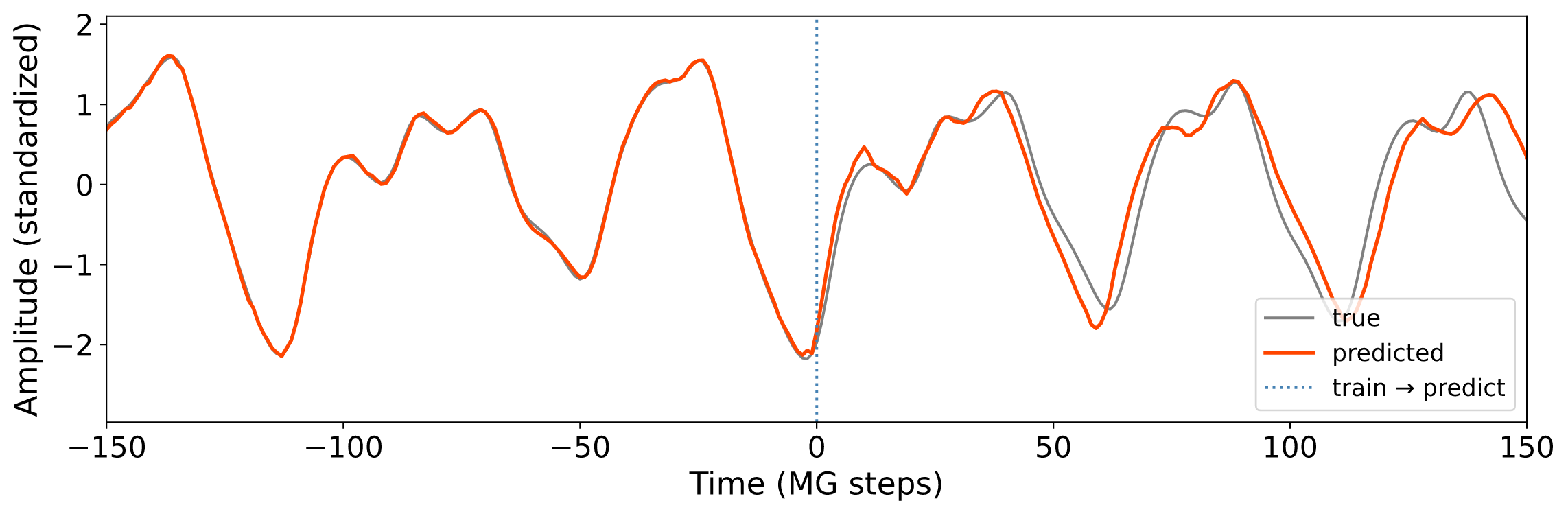}
        \caption{$\tau = 18$}
        \label{fig:nonvn_time_18}
    \end{subfigure}
    
    \vspace{1em}
    
    \begin{subfigure}{\textwidth}
        \centering
        \includegraphics[width=0.75\textwidth]{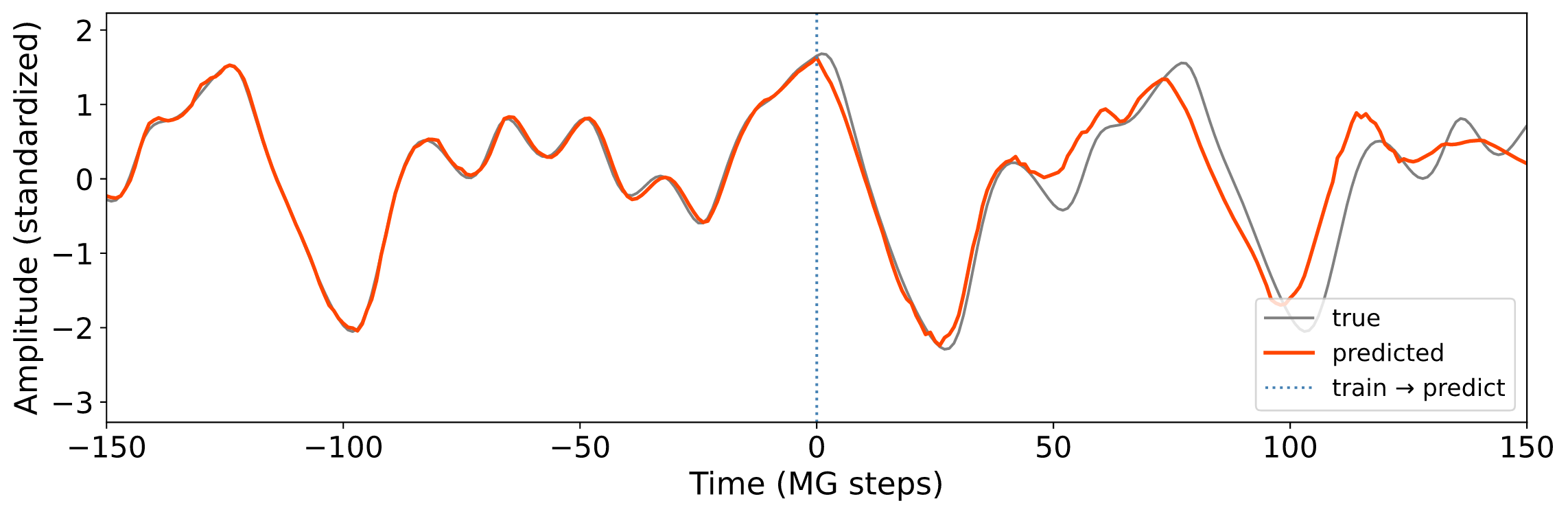}
        \caption{$\tau = 21$}
        \label{fig:nonvn_time_21}
    \end{subfigure}
    
    \caption{Autonomous prediction of the 
    Mackey-Glass time series using the non-VN method. 
    Top: $\tau = 18$; bottom: $\tau = 21$. The vertical dashed line marks the 
    transition from teacher-forced training to closed-loop 
    autonomous prediction.}
    \label{fig:nonvn_timeseries}
\end{figure}

Figure~\ref{fig:nonvn_timeseries} shows representative time-series for the
non-VN approach. The reservoir captures the broad 
quasi-periodic structure of the MG signal in both 
delay regimes, though with larger amplitude deviations 
and earlier phase divergence than the VN approach. The 
performance gap is consistent with the difference in 
effective feature dimensionality between the two 
approaches: virtual nodes expand the readout by sampling 
the reservoir state at multiple sub-timestep intervals, 
whereas the non-VN approach extracts all predictive 
information from a single per-timestep snapshot of node 
voltages. This suggests that the diversity of 
readouts across the 500-node network, while rich, does 
not fully substitute for the temporal feature expansion 
provided by virtual nodes at this network scale. 
However, physical nanowire network reservoirs can readily scale to 
orders of magnitude more nodes~\cite{diaz-alvarezEmergent2019,Lilak2021,Milano_2022,zhu2023online}, and it is plausible that increasing 
the number of physical nodes would expand the  
feature diversity and expressiveness sufficiently to close this performance 
gap without recourse to virtual node augmentation. 
Exploring this scaling behaviour is left for future work.

\subsection{Long-Horizon Prediction Analysis}
\label{sec:longhorizon}

Autonomous prediction is extended to $T = 500$ timesteps 
to assess long-horizon behaviour. As point-wise accuracy 
is expected to degrade for any chaotic system due to 
exponential error amplification, the analysis focuses on 
how well the predicted trajectories preserve the phase 
dynamics and spectral properties of the true MG dynamics, 
rather than tracking the true signal step-by-step.

\paragraph{Phase Space.}

\begin{figure}[htbp]
    \centering
    \begin{subfigure}[b]{0.45\textwidth}
        \centering
        \includegraphics[width=\linewidth]{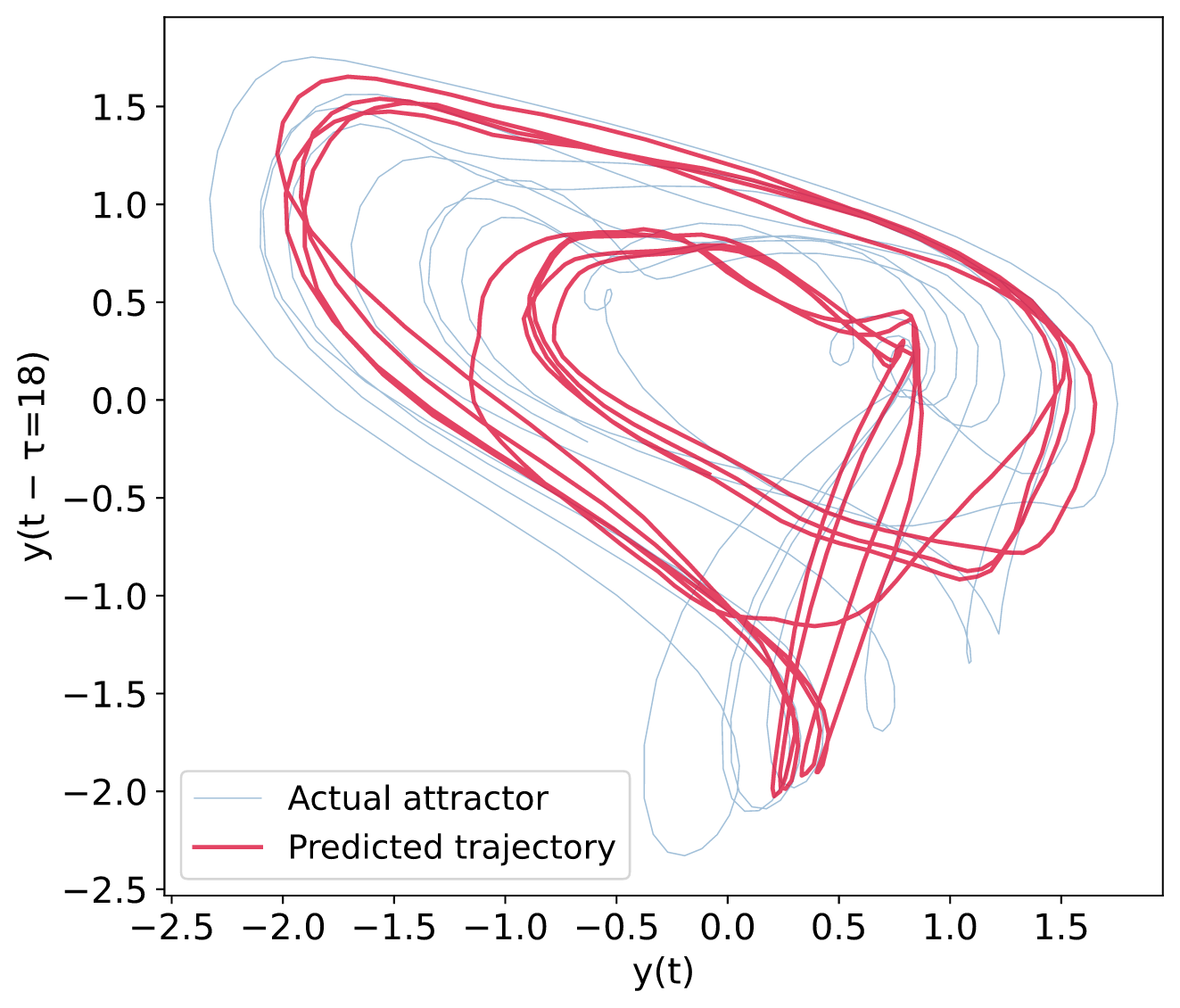}
        \caption{$\tau = 18$}
        \label{fig:vn_phase_18}
    \end{subfigure}
    \hfill
    \begin{subfigure}[b]{0.45\textwidth}
        \centering
        \includegraphics[width=\linewidth]{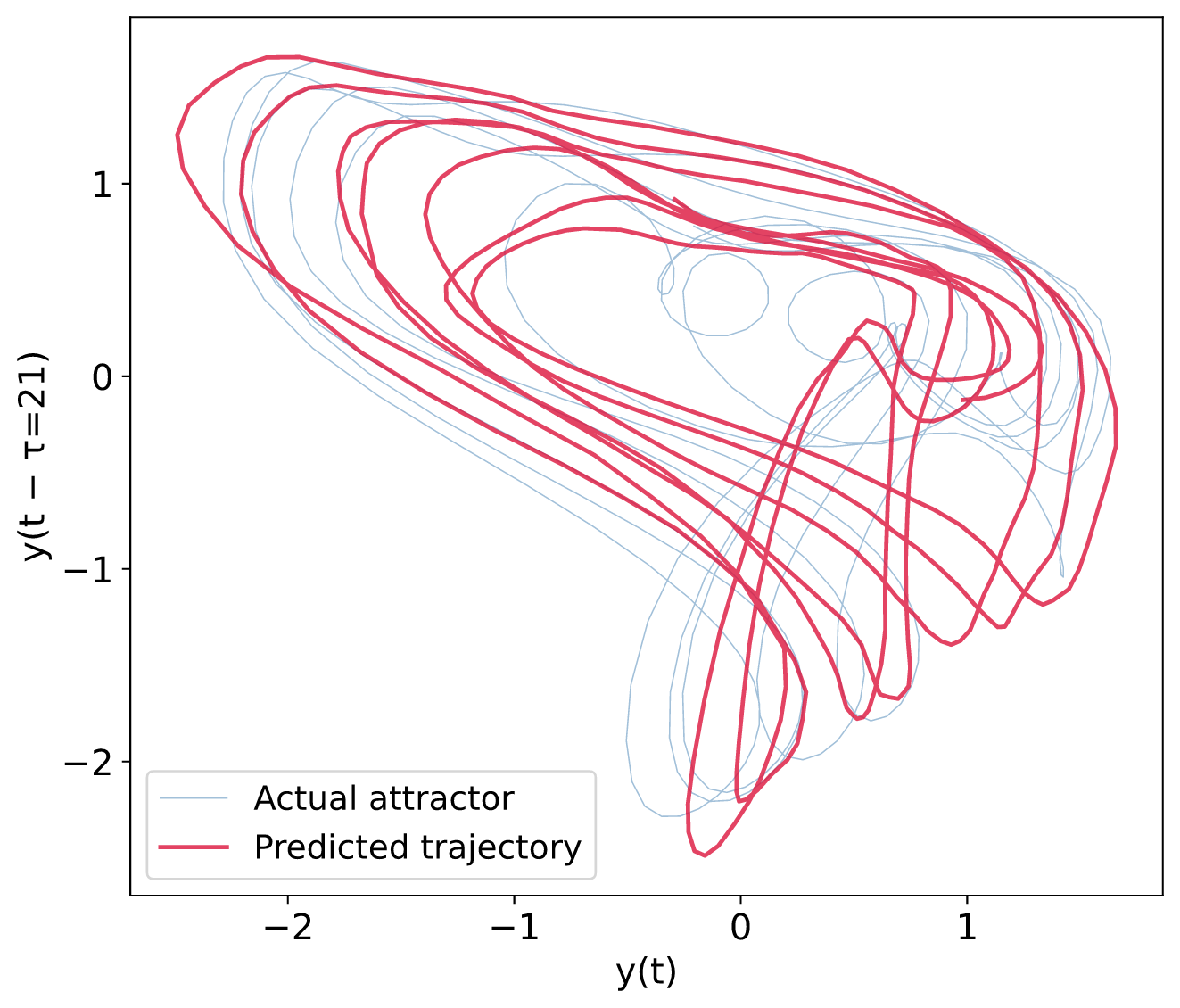}
        \caption{$\tau = 21$}
        \label{fig:vn_phase_21}
    \end{subfigure}
    \caption{Reconstructed phase space attractors 
    over $T = 500$ timesteps using the VN method for (a)
    $\tau = 18$ and (b) $\tau = 21$.}
    \label{fig:vn_phase}
\end{figure}

\begin{figure}[htbp]
    \centering
    \begin{subfigure}[b]{0.45\textwidth}
        \centering
        \includegraphics[width=\linewidth]{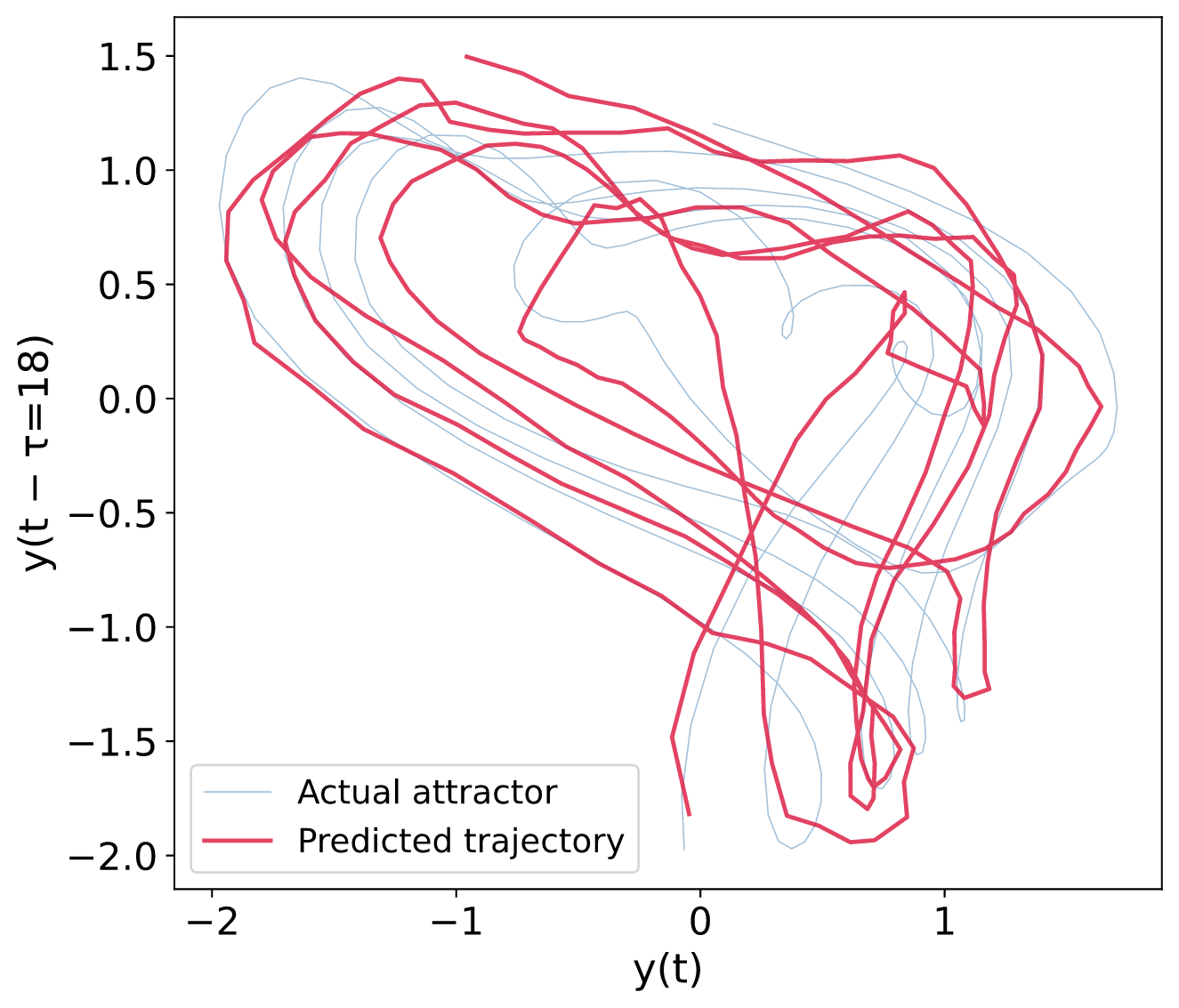}
        \caption{$\tau = 18$}
        \label{fig:nonvn_phase_18}
    \end{subfigure}
    \hfill
    \begin{subfigure}[b]{0.45\textwidth}
        \centering
        \includegraphics[width=\linewidth]{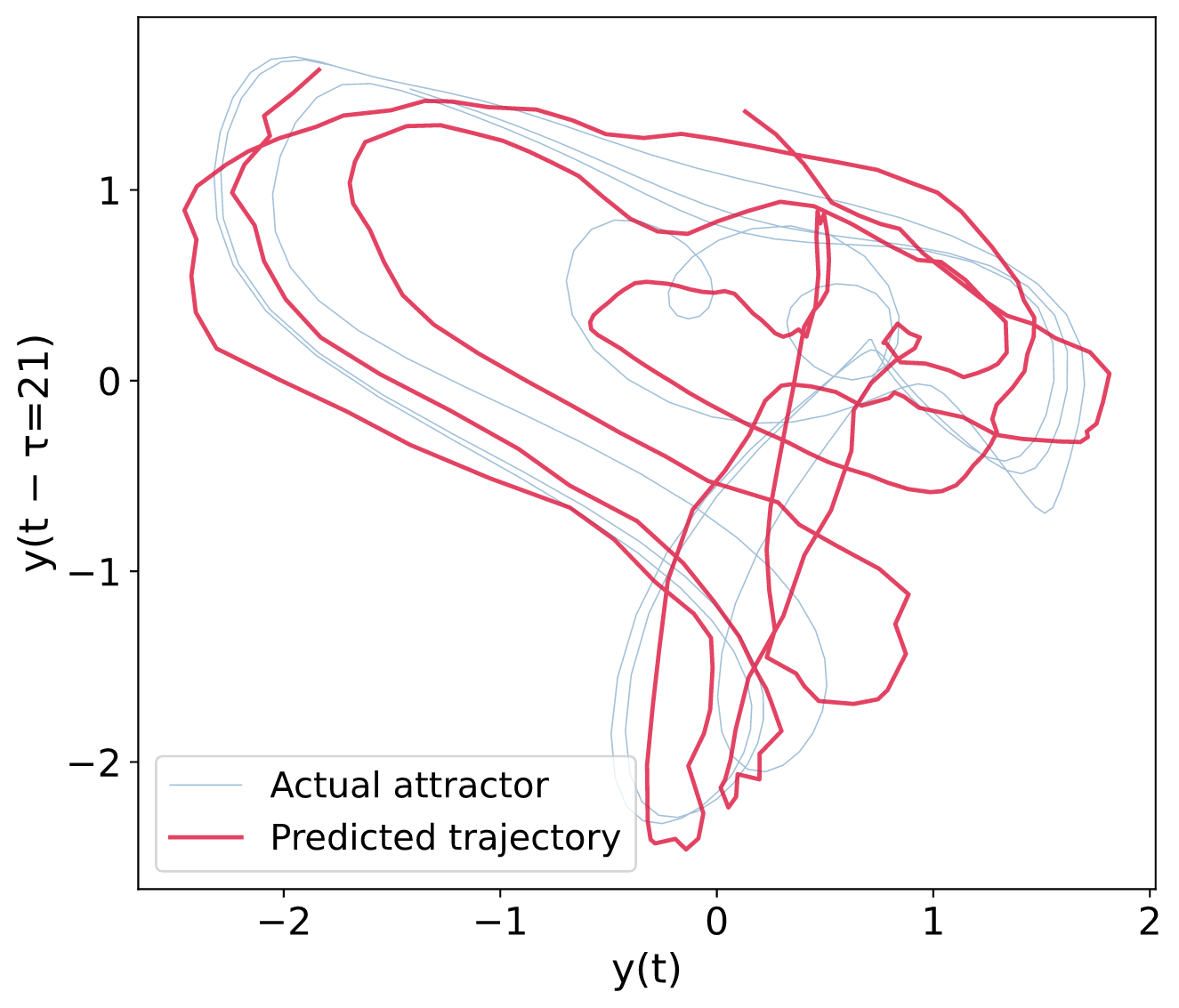}
        \caption{$\tau = 21$}
        \label{fig:nonvn_phase_21}
    \end{subfigure}
    \caption{Reconstructed phase space attractors 
    over $T = 500$ timesteps using the non-VN method for 
    (a) $\tau = 18$ and (b) $\tau = 21$.}
    \label{fig:nonvn_phase}
\end{figure}

Figures~\ref{fig:vn_phase} and~\ref{fig:nonvn_phase} 
show the reconstructed phase space attractors for the 
predicted trajectories using the VN and non-VN methods, 
respectively. In both cases, the predicted trajectory 
broadly follows the structure of the true MG attractor, 
reproducing the approximate phase space extent and the 
qualitative folding behaviour at both $\tau$ values.
The predicted trajectories remain bounded throughout the 
500-timestep window in both approaches, suggesting that 
the closed-loop feedback does not cause the prediction 
to diverge catastrophically at longer horizons.

The predicted attractor for the VN method follows the 
true trajectory more closely, with loops that are more 
tightly bounded and consistently wound. The non-VN 
predicted attractor shows greater dispersion, with 
individual loops deviating more broadly from the true 
attractor boundary, particularly at $\tau = 21$. This 
is consistent with the lower short-term accuracy of the 
non-VN approach and the compounding of prediction errors 
over longer horizons. At $\tau = 21$, both approaches 
produce attractors that are geometrically more extended 
than at $\tau = 18$, reflecting the higher effective 
dimensionality of the longer-delay dynamics, and both 
predicted attractors scale accordingly in extent.

\clearpage

\paragraph{Power Spectral Density.}

\begin{figure}[htbp]
    \centering
    \begin{subfigure}[b]{0.45\textwidth}
        \centering
        \includegraphics[width=\linewidth]{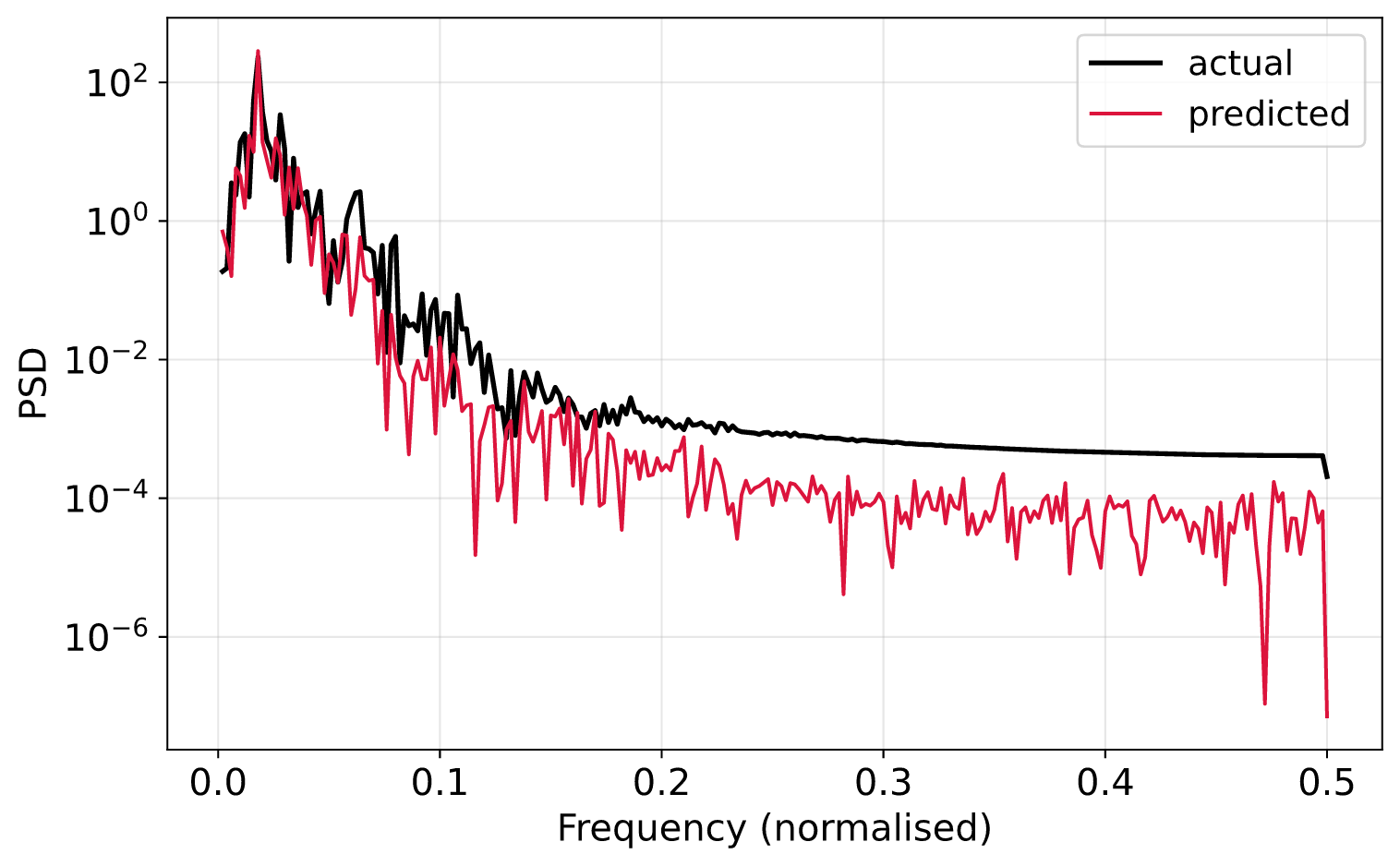}
        \caption{$\tau = 18$}
        \label{fig:vn_psd_18}
    \end{subfigure}
    \hfill
    \begin{subfigure}[b]{0.45\textwidth}
        \centering
        \includegraphics[width=\linewidth]{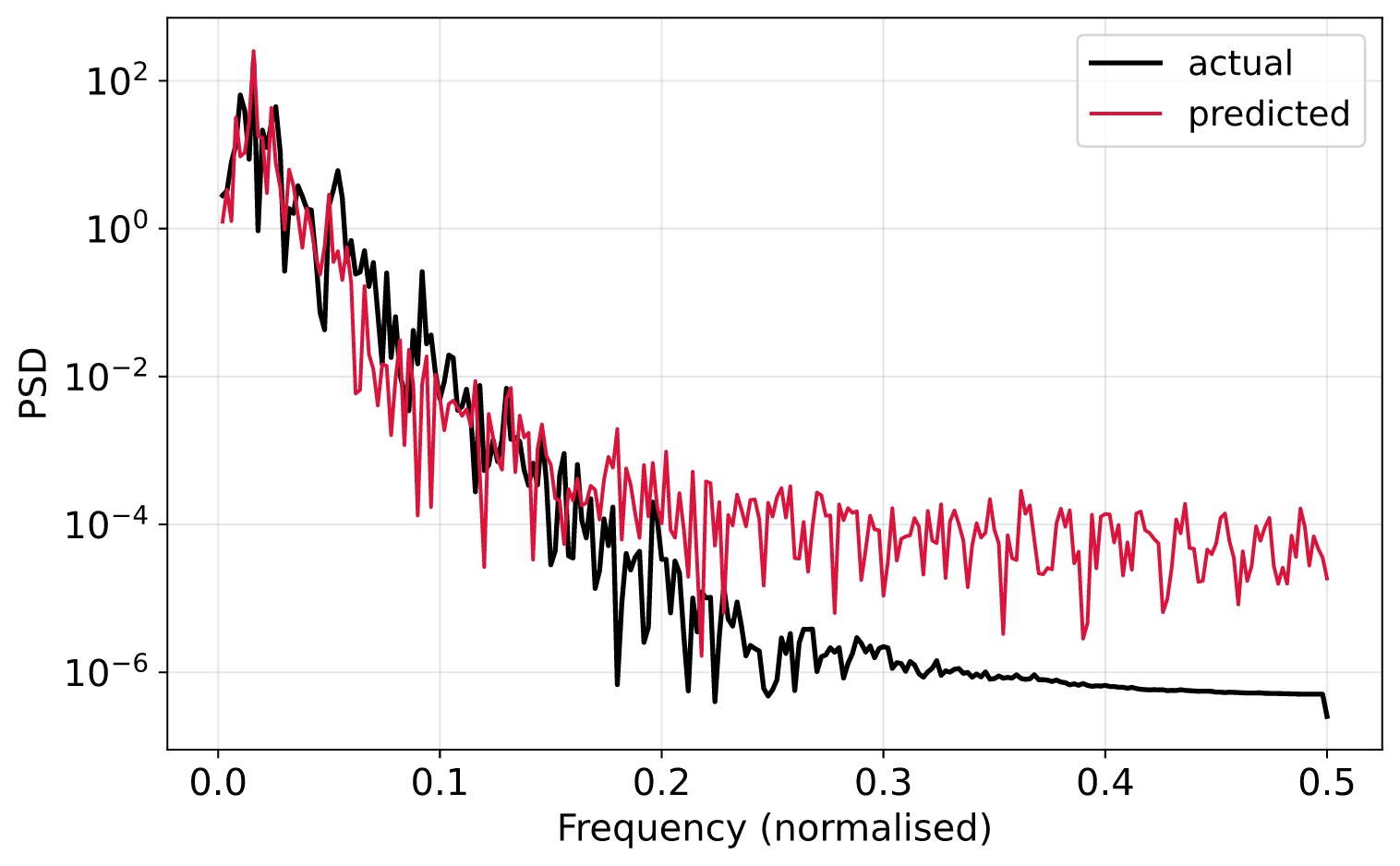}
        \caption{$\tau = 21$}
        \label{fig:vn_psd_21}
    \end{subfigure}
    \caption{Normalised power spectral density (PSD) of 
    predicted and true Mackey-Glass signals over 
    $T = 500$ timesteps using the VN method for 
    $\tau = 18$ (left) and $\tau = 21$ (right).}
    \label{fig:vn_psd}
\end{figure}

\begin{figure}[htbp]
    \centering
    \begin{subfigure}[b]{0.45\textwidth}
        \centering
        \includegraphics[width=\linewidth]{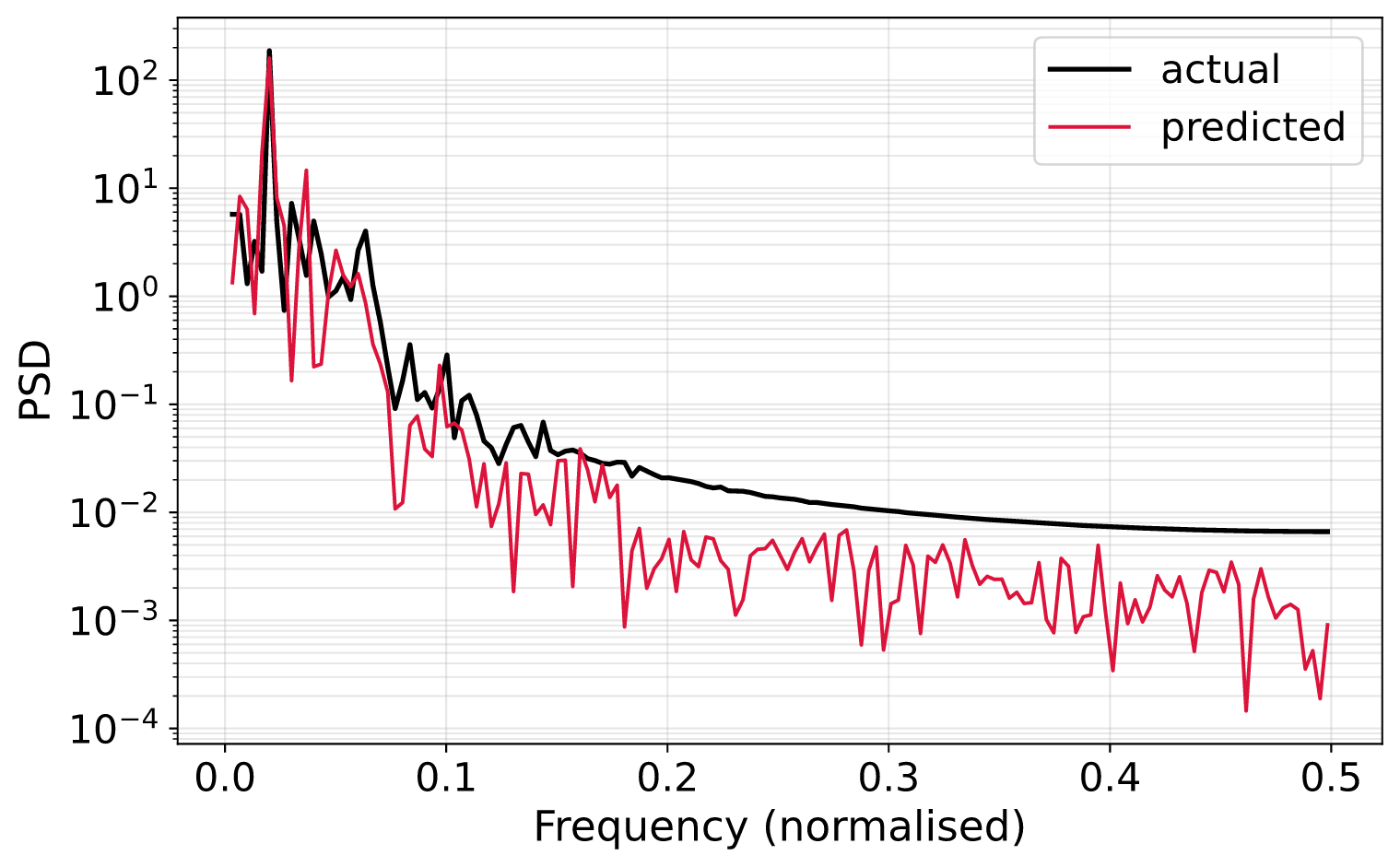}
        \caption{$\tau = 18$}
        \label{fig:nonvn_psd_18}
    \end{subfigure}
    \hfill
    \begin{subfigure}[b]{0.45\textwidth}
        \centering
        \includegraphics[width=\linewidth]{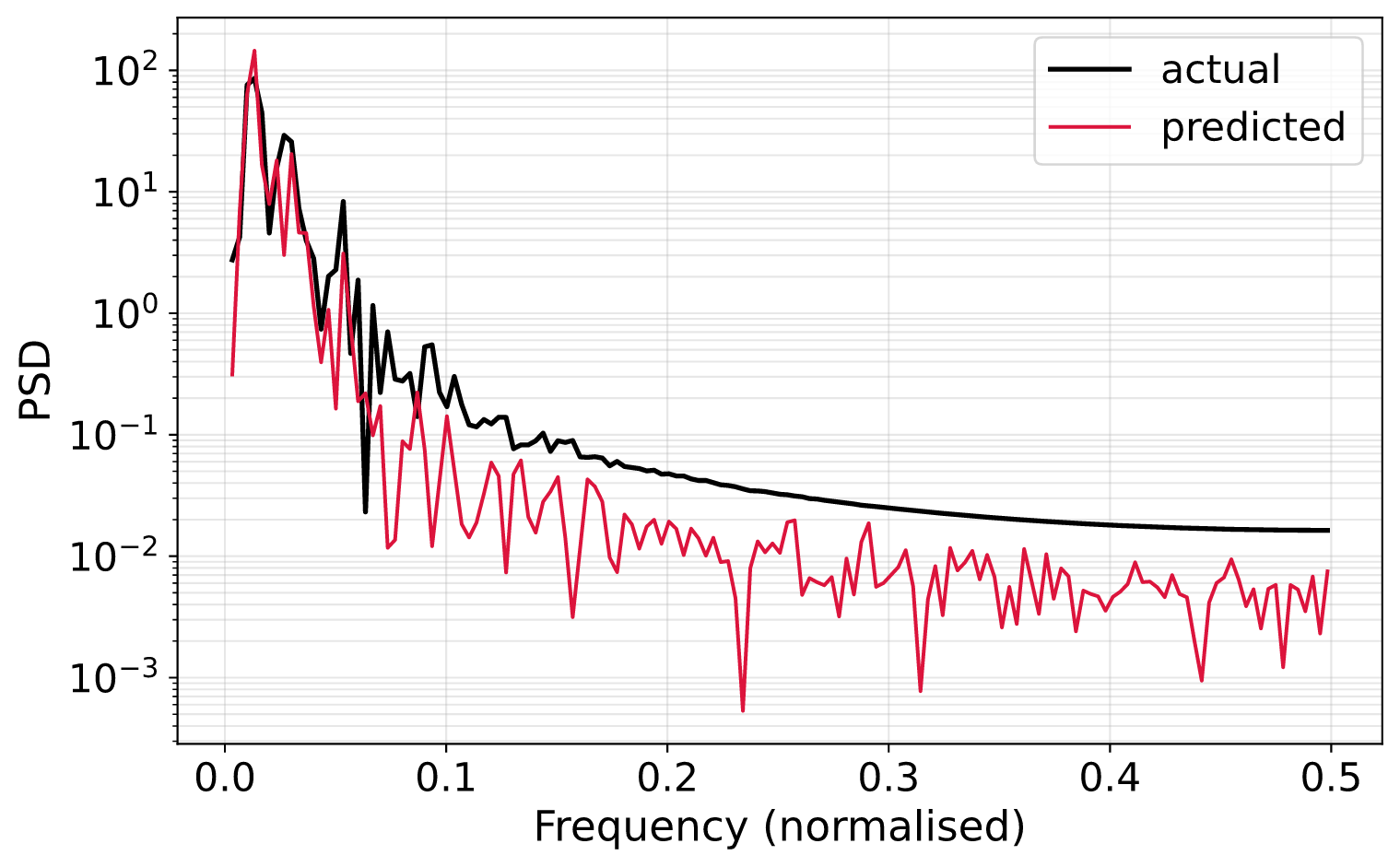}
        \caption{$\tau = 21$}
        \label{fig:nonvn_psd_21}
    \end{subfigure}
    \caption{Normalised power spectral density (PSD) of 
    predicted and true Mackey-Glass signals over 
    $T = 500$ timesteps using the non-VN method for 
    $\tau = 18$ (left) and $\tau = 21$ (right). Layout 
    and conventions identical to Fig.~\ref{fig:vn_psd}.}
    \label{fig:nonvn_psd}
\end{figure}

Figures~\ref{fig:vn_psd} and~\ref{fig:nonvn_psd} compare 
the normalised PSDs of the predicted and true signals 
over 500 timesteps for the VN and non-VN approaches 
respectively. In both cases, the dominant low-frequency 
spectral content of the true MG signal is broadly, but 
not precisely, reproduced at both $\tau$ values. The 
primary spectral peak and the general low-frequency 
power envelope are captured reasonably well, suggesting 
that the predicted signal retains the principal frequency 
components of the true dynamics at long horizons.

At higher frequencies, the PSD for both approaches 
deviates from that of the true signal, consistent with 
the accumulation of prediction errors over 500 timesteps.
For $\tau = 18$, both approaches underpredict power in 
the high-frequency signal components, while for 
$\tau = 21$, the VN method overpredicts at high 
frequencies and the non-VN method underpredicts. 
Overall, the PSD results are consistent with the phase 
space analysis: both approaches capture the dominant 
dynamical structure of the MG signal over long horizons, 
with the VN approach providing a closer spectral match 
and the non-VN approach capturing dominant frequencies 
while showing more pronounced high-frequency 
discrepancies.
\clearpage

To quantify the PSD results, we computed the Jensen--Shannon (JS) distance between the normalized PSDs of predicted and true trajectories, and NRMSE at the full $T = 500$ prediction horizon. The JS distance is defined as the square root of the JS divergence,
\begin{equation}
    \mathrm{JSD}(P \,\|\, Q) = \tfrac{1}{2} D_{\mathrm{KL}}(P \,\|\, M) + \tfrac{1}{2} D_{\mathrm{KL}}(Q \,\|\, M), \qquad M = \tfrac{1}{2}(P + Q),
\end{equation}
where $D_{\mathrm{KL}}$ denotes the Kullback--Leibler divergence and $P$, $Q$ are the (normalized) PSDs of the predicted and true trajectories, respectively; i.e.\ $\mathrm{JS distance} = \sqrt{\mathrm{JSD}(P\,\|\,Q)}$. Unlike the divergence itself, the JS distance satisfies the triangle inequality and is thus a proper metric on probability distributions.

For the non-VN approach across the 10 evaluation trials (500-node network as described in Sec.~\ref{sec:methods}): at $\tau = 21$, JS distance was $0.271 \pm 0.104$ and NRMSE was $0.226 \pm 0.022$; at $\tau = 18$, JS distance was $0.172 \pm 0.056$ and NRMSE was $0.174 \pm 0.032$.  For the VN approach: at $\tau = 21$, JS distance was $0.254 \pm 0.110$ and NRMSE was $0.206 \pm 0.036$, while at $\tau = 18$, JS distance was $0.226 \pm 0.100$ and NRMSE was $0.172 \pm 0.092$. The closer spectral match and lower error at $\tau = 18$ relative to $\tau = 21$ is consistent with the qualitative phase-space and PSD comparisons above.

Taken together, the short-term and long-horizon results 
highlight three aspects of this work. First, fully 
autonomous closed-loop nanowire network PRC is shown 
to extend to $\tau = 21$, a delay regime not previously 
evaluated for this class of physical reservoir computers. Second, 
the non-VN strategy achieves autonomous 
Mackey-Glass prediction without virtual node 
augmentation, providing a simpler baseline that, while less accurate 
than the VN approach, captures the qualitative dynamical 
structure of the signal. Third, the long-horizon analysis 
shows that both approaches produce bounded, 
attractor-consistent trajectories over 500 timesteps. 
A more rigorous characterization of long-horizon 
statistical fidelity remains an avenue for future work.

\section{Conclusion}
\label{sec:conclusion}

This study investigated autonomous closed-loop prediction 
of the Mackey--Glass chaotic time series using a 
simulated neuromorphic nanowire network as a physical 
reservoir, comparing two readout strategies across two 
values of the MG time delay parameter.

The VN approach, using Bayesian-optimised readout nodes 
with 20 virtual nodes each, achieved autonomous 
prediction accuracies of $0.904 \pm 0.039$ and 
$0.897 \pm 0.033$ at $\tau = 18$ and $\tau = 21$ 
respectively over $T = 100$ timesteps, consistent with 
prior nanowire network PRC results at $\tau = 18$ and 
extending autonomous prediction to the more complex 
$\tau = 21$ regime. The non-VN approach, which uses all 
physical node readouts without temporal multiplexing, 
achieved lower but still meaningful accuracies of 
$0.815 \pm 0.059$ and $0.762 \pm 0.069$ at the same 
$\tau$ values, without any virtual node construction 
or node selection optimisation.

Long-horizon analysis over $T = 500$ timesteps showed 
that both approaches produce bounded trajectories that 
broadly preserve the qualitative attractor structure 
and dominant spectral content of the true MG signal, 
though with greater dispersion in the non-VN case, 
particularly at $\tau = 21$. The performance gap between 
the two approaches is attributable to the difference in 
effective feature dimensionality, and it is plausible 
that scaling to the larger physical node counts 
~\cite{diaz-alvarezEmergent2019,Milano_2022} could reduce this gap without recourse to 
virtual node augmentation. As this study uses simulated 
networks, extrapolation to physically fabricated 
large-scale arrays remains to be validated 
experimentally.

Taken together, these results demonstrate that 
neuromorphic nanowire networks can support fully 
autonomous chaotic time series prediction across 
multiple dynamical regimes, and that the intrinsic 
spatial diversity of network readouts alone carries 
meaningful predictive information. Future work should 
investigate the effect of network size on non-VN 
prediction performance, explore principled readout 
selection strategies for the non-VN case, and extend 
evaluation to longer prediction horizons and additional 
chaotic benchmarks.

\begin{credits}
\subsubsection{\ackname} The authors acknowledge the use of the NCI high performance computing facility supported by the University of Sydney. Y.X. is supported by an Australian Government Research Training Program (RTP) scholarship.

\subsubsection{\discintname}
Z.K. owns stock in Emergentia, Inc.
\end{credits}
%
%
%
\bibliographystyle{splncs04}
\bibliography{refs}
\end{document}